\documentclass[letterpaper,prl,twocolumn,showpacs,superscriptaddress]{revtex4} 
\usepackage[mathscr]{eucal}
\usepackage{amsmath}
\usepackage{amssymb}
\usepackage{amsfonts}
\usepackage{ae}
\usepackage{epsfig}
\usepackage{subfigure}
\usepackage{caption}
\usepackage{graphicx}
\usepackage{dcolumn}
\usepackage{bm}
\usepackage[latin1]{inputenc}
\usepackage{color}

\begin{document}
\title{Multimode non-classical light generation through the OPO
 threshold}

\author{B. Chalopin}\email{chalopin@spectro.jussieu.fr}\affiliation{Laboratoire
 Kastler Brossel, Universit\'e Pierre et Marie-Curie-Paris 6, ENS,
 CNRS ; 4 place Jussieu, 75005 Paris, France} \author{F. Scazza}
\affiliation{Laboratoire Kastler Brossel, Universit\'e Pierre et
 Marie-Curie-Paris 6, ENS, CNRS ; 4 place Jussieu, 75005 Paris,
 France} \author{C. Fabre} \affiliation{Laboratoire Kastler Brossel,
 Universit\'e Pierre et Marie-Curie-Paris 6, ENS, CNRS ; 4 place
 Jussieu, 75005 Paris, France}
\author{N. Treps}\affiliation{Laboratoire Kastler Brossel,
 Universit\'e Pierre et Marie-Curie-Paris 6, ENS, CNRS ; 4 place
 Jussieu, 75005 Paris, France}

\date{\today}
\begin{abstract}
 We show that an Optical Parametric Oscillator which is
 simultaneously resonant for several modes, either spatial or
 temporal, generates both below and above threshold a multimode
 non-classical state of light consisting of squeezed vacuum states in
 all the non-oscillating modes. We
 confirm this prediction by an experiment dealing with the degenerate
 TEM$_{01}$ and TEM$_{10}$ modes. We show the
 conservation of non-classical properties when the threshold is
 crossed. The
 experiment is made possible by the implementation of a new method to
 lock the relative phase of the pump and the injected beam.
\end{abstract}

\maketitle
Optical parametric oscillators are among the best generators of
non-classical states of light. Below the oscillation threshold, they
have been shown to generate squeezed vacuum states \cite{KimbleWu}
and bi-partite entangled states. Above the
oscillation threshold, they generate intensity correlated twin beams
\cite{Heidmann}, squeezed reflected pump \cite{Kasai}, bright bi-partite
\cite{Peng} or tri-partite \cite{Villar} entangled states. Very
impressive amounts of squeezing (11dB) have been recently observed
below threshold\cite{Schnabel}. Experimental results are less spectacular above
threshold, because of the detrimental effect of the pump beam excess
noise. Generally speaking, the non-classical properties increase when
one approaches from below or from above the oscillation threshold, or
any other bifurcation point of the non-linear dynamics of the
device\cite{Solvay,Gigi}, but it is not always the case: for instance the signal-idler intensity difference
squeezing is independent of the pumping level. This property of
"non-critially squeezed light" has been further explored by the
Valencia group\cite{Navarrete} in the context of quantum imaging,
which has found other squeezing effects that are independent of the
pumping level. These effects are related to spatial symmetry-breaking,
either translational\cite{Valcarcel1} or rotational\cite{Navarrete}: in a
cavity of cylindrical symmetry, the parametric gain is the same for
Gaussian modes TEM$_{01}$ where the two lobes are aligned along any
direction in the transverse plane. However, above the oscillation
threshold, a unique TEM$_{01}$ mode is produced. It is shown that
this spontaneous symmetry breaking "induces" the generation of a
squeezed vacuum state in the mode orthogonal to the emitted one with a
squeezed value independent of the pumping level and clamped to its
value at threshold. The emitted field is thus a two-mode non-classical
field, made of the superposition of a bright mode and a
vacuum-squeezed mode.

In this paper, we show theoretically that this important result
can be generalized to a much wider class of situations, involving
either spatial or temporal modes, and we check experimentally that one
can efficiently produce in this way multimode non-classical light.

Let us envision the situation in which the cavity of the OPO is
simultaneously resonant on several modes, which can be either spatial
modes (Hermite-Gauss modes or more complicated patterns in transverse
degenerate cavities), or frequency modes (separated by the free
spectral range of the cavity). The annihilation operators $\hat a_{\ell}$ associated with
these different modes and the pump mode operator $\hat b$ obey then
the following well-known evolution equations, describing the effect of
the parametric splitting of pump photons into couples of signal and
idler photons respectively in modes $\ell$ and
$\ell'$ \begin{equation} \tau\frac{d}{dt}\hat a_{\ell}=-\gamma \hat
 a_{\ell} + \sum_{\ell'}G_{\ell,\ell'}\hat b \hat
 a_{\ell'}^{\dagger}+\sqrt{2 \gamma} \hat a_{\ell}^{in}
\end{equation}
assuming equal cavity losses $\gamma$ for all the modes $\ell$. $\tau$
is the cavity round-trip time. The pump is described by a single mode
$\hat b$ having a well-defined spatial and temporal variation.  $ \hat
a_{\ell}^{in}$ are the input modes. Like in \cite{Navarrete2}, we take
into account the fact that the parametric coupling coefficients $G_{\ell,\ell'}$ between the signal modes and
the pump vary according to the strength of the overlap between
the modes $\hat a_{\ell}$, $\hat a_{\ell'}$ and
$\hat b$. $G_{\ell,\ell'}$ is a symmetric matrix which can always be diagonalized: let us call
$\Lambda_k$ its real eigenvalues. The corresponding eigenvectors are
"supermodes"\cite{Patera1,Patera2,Lopez}, associated to annihilation
operators $\hat S_k$, which obey the following decoupled equations:
\begin{equation}
 \tau\frac{d}{dt}\hat S_k = -\gamma \hat S_k + \Lambda_k \hat b \hat S_k^{\dagger}+\sqrt{2 \gamma} \hat S_k^{in}
\end{equation}
The mean intensity of the different supermodes $\hat S_k$ is zero as
long as the system stays below threshold. Let us call $\hat S_1$ the
supermode associated to the eigenvalue $\Lambda_1$ of highest modulus
(in the situation studied in \cite{Navarrete}, this eigenvalue is
degenerate, because of symmetries in the considered device). When one
increases the pump power, this mode will reach first the oscillation
threshold. It is easy to show that below this threshold all these
modes are in squeezed vacuum states, the squeezing increasing when one
approaches the threshold. At threshold, the noise on the squeezed
quadrature has a variance at zero noise frequency $V_{k, min}$
(normalized to vacuum noise) equal to\cite{Patera1}:
\begin{equation}\label{squeezing}
 V_{k, min}=\left(\frac{|\Lambda_k|-|\Lambda_1|}{|\Lambda_k|+|\Lambda_1|}\right)^2
\end{equation}
meaning that all modes with eigenvalues equal to $\Lambda_1$ or
$-\Lambda_1$ are perfectly squeezed. This property has been recently
checked experimentally in the case of a cavity with cylindrical
symmetry by two groups\cite{Lassen2009,Janousek}, which have
demonstrated simultaneous squeezing on two orthogonal first-order
hermite gaussian modes TEM$_{10}$ and TEM$_{01}$ produced by a
degenerate OPO pumped by a TEM$_{00}$ below the oscillation threshold.

Let us now consider the above threshold case, but close enough to the
threshold so that one can neglect the distortion of the pump mode
shape inside the crystal due to pump depletion. The first supermode
$k=1$ oscillates and the intracavity pump mode has a nonzero mean
value $\left\langle \hat b\right\rangle$ "clamped" at a value
$\gamma/\Lambda_1$, or $- \gamma/\Lambda_1$, independent of the pump
input intensity. The others are still below threshold and have zero
mean values. The evolution equations of these modes are obtained by
the usual procedure of linearization of operators equations around the
mean values. One gets:
\begin{equation}\label{above6}
 \tau \frac d{dt} \hat{S}_k =-\gamma \hat{S}_k \pm \gamma \frac{\Lambda_k}{\Lambda_1} \hat{S}_k^{\dagger} + \sqrt{2 \gamma }\hat{S}_{in,k} \quad (k\neq 1)
\end{equation}
from which one easily derives that all these modes are also "clamped"
to the squeezed vacuum state that they had reached when approaching
the threshold from below, as long as pump depletion does not distort
significantly the pump mode shape. Their minimum variance is then
given by Eq.(\ref{squeezing}) whatever the pump power. All modes for
which $|\Lambda_k| \simeq |\Lambda_1|$ are therefore significantly
squeezed. We have thus shown that {\it an OPO simultaneously resonant
 on different modes produces above threshold a multimode nonclassical
 state consisting of several squeezed vacuum superposed to a bright
 mode}.

\begin{figure}%
 \includegraphics[width=\columnwidth]{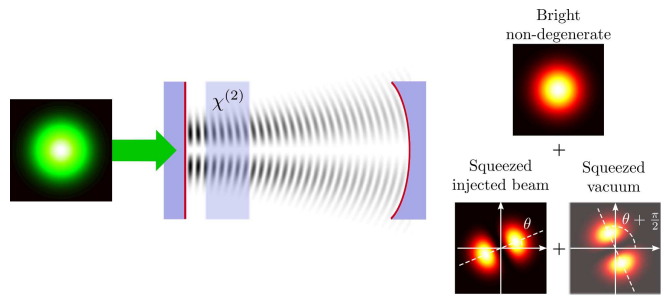}%
 \caption{Principle of the squeezing generation in a degenerate OPO
   above threshold}%
 \label{principe}%
\end{figure}

The principle of our experiment is shown on Fig.
\ref{principe}. The OPO cavity is pumped by a TEM$_{00}$ mode and the cavity is
simultaneously resonant for the two transverse modes TEM$_{10}$
and TEM$_{01}$. The mode matching properties of our device are such that
the lowest oscillation threshold is obtained for a couple of non frequency degenerate signal and idler modes both in a transverse TEM$_{00}$ mode.
Consequently, the two frequency degenerate
EM$_{10}$
and TEM$_{01}$ transverse modes, having a higher threshold, should be in a squeezed vacuum state, both below and above the oscillation threshold.
This is what we want to check in the experiment

\begin{figure}[h]%
 \includegraphics[width=\columnwidth]{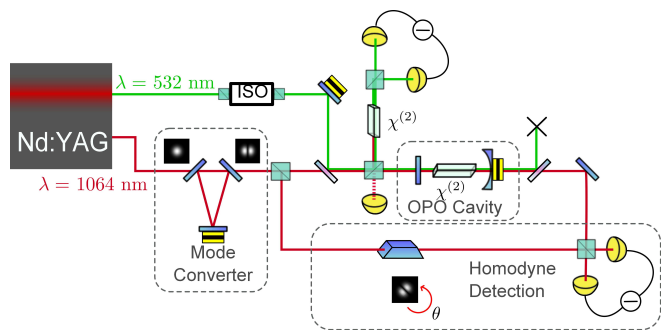}%
 \caption{Schematics of the experimental setup}%
 \label{expestup}%
\end{figure}

The experimental setup is depicted on Fig. \ref{expestup}. We build a
two-mode degenerate OPO exploiting the simultaneous resonance of both
TEM$_{10}$ and TEM$_{01}$ in a linear cavity, and we placed a
$1mm\times2mm\times10mm $PPKTP non-linear
crystal inside. The high non-linear
efficiency of the PPKTP enables us to use a single-pass $532nm$
pump beam from a frequency-doubled $1064nm$ YAG laser. The
input coupler is a highly reflective plane mirror at $1064nm$
with $R=99.8\%$ and the output coupler is a spherical mirror
of radius of curvature $50mm$ and reflectivity $R=98.3\%$. The
intra-cavity losses are around $0.2\%$ and are mainly due to the
non-linear crystal. The cavity finesse is $\mathcal{F}=$ 300 with an
escape efficiency of $80\%$. This value is a trade-off between
the level of squeezing we can observe with this OPO and the power of
the pump at the threshold of the OPO. The cavity length is $47mm$
and results in a bandwidth of $11MHz$.

We generate a horizontal TEM$_{10}$ mode with a mode converter cavity
(MC) seeded with a misaligned TEM$_{00}$. This mode is seeded in the
OPO cavity to achieve alignment and locking at resonance. The pump is
a TEM$_{00}$ mode. Its mode-matching is a delicate operation: as
reported in \cite{Lassen2006}, the optimal pump profile for the
amplification of a TEM$_{10}$ and a TEM$_{01}$ would be a combination
of TEM$_{00}$, TEM$_{20}$ and TEM$_{02}$. For simplicity reasons, we
chose to use a TEM$_{00}$ mode only, whose waist is ajusted to
maximize the amplification gain of both the TEM$_{10}$ and the
TEM$_{01}$. The degeneracy of the cavity for these two modes is easy
to obtain when the cavity is empty. The periodic poling of the PPKTP
crystal induces a slight disymmetry, which makes the cavity
non-degenerate, but can be compensated with a fine tuning of the
crystal temperature.
\begin{figure}%
 \includegraphics[width=\columnwidth]{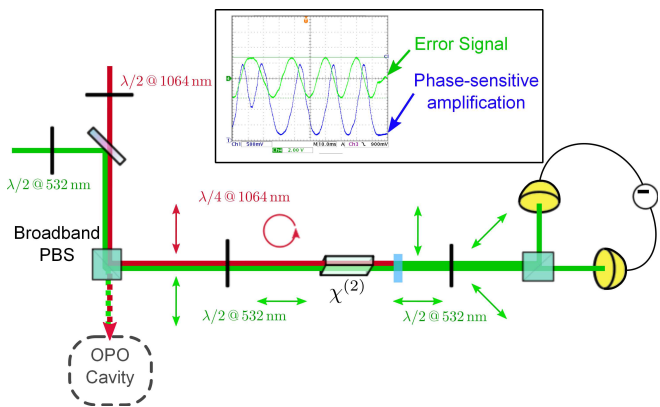}%
 \caption{Error signal generation for the relative phase between OPO
   seed and pump beams. Polarisations of the beams, as shown on the
   figure, are choosen to allow interferences only on the last
   beamsplitter. The phase-sensitive
   amplification and the error signal (windowed) show
   perfect correlation.}
 \label{reativephase}%
\end{figure}

To keep the OPO stable while crossing the oscillation threshold, and
continuously compare the two regimes, one must perform all the
necessary lockings on the beams upstream from the cavity. The cavity
length is locked at resonance using the Pound-Drever-Hall technique
\cite{PDH}. When the cavity is locked on a TEM$_{00}$ resonance, the
pump threshold is $250mW$, while for the TEM$_{10}$ resonance the
threshold becomes $450mW$. The relative phase between the seed
and the pump of the OPO has to be locked in the de-amplification regime
in order to observe amplitude squeezing. To this end, an error signal
is generated independently of the OPO with a technique depicted on Fig
\ref{reativephase}. An interferometer is built between the input pump
and the seed frequency doubled within a PPKTP crystal. Since the optical path is identical
for all the modes, the interference between the pump and the frequency
doubled seed, that depends on the two intensities and the relative phase
$\phi$ between the pump beam and the seed of the OPO, can be used
as an error
signal.
\begin{eqnarray}
 s(\phi)\propto \sqrt{I_\text{pump}} I_\text{seed} \cos (2\phi - \phi_0)
 \label{eq:errorsignal}
\end{eqnarray}
where $\phi_0$ is an offset phase, that can be tuned simply by moving
the crystal longitudinally in order to lock the relative phase
$\phi$. The key points for this non-linear interferometer are on the one hand to use a
broadband polarizing beamsplitter which enables us to independently
tune the powers of the two beams sent in the interferometer, and on
the other hand the high non-linear coefficient of the PPKTP that makes
possible the single pass frequency doubling of an infrared beam with a
sufficient efficiency. The locking turned out to be very stable
once the two relative powers sent inside the interferometers have
been carefully chosen. In our case, we sent about $30mW$ of
infrared power and less than $1mW$ of green
power.

The quantum state of the output modes of the OPO is analyzed with a
homodyne detection. The local oscillator is a TEM$_{10}$ of arbitrary
orientation, generated by rotating the mode transmitted by the MC with
a Dove prism \cite{Janousek}. It is first aligned when the mode
converter is locked on the TEM$_{00}$ with a visibility above
$98\%$, so that the orthogonality of the eigenmodes of the MC
assures the orthogonality of the ones measured with the homodyne
detection.
Below the threshold, the output of the OPO shows multimode squeezing
on both the injected TEM$_{10}$ and the vacuum TEM$_{01}$ modes with
$20\%$ noise reduction ($1dB$). The system has more losses
and smaller bandwidth than OPOs optimized for squeezing which explains
the low amount of squeezing. This value has to be compared with the
squeezing observed in the same experimental conditions on a TEM$_{00}$
mode, which is $1.5dB$. The ratio between squeezing on these two
modes is in agreement with the prediction of Eq.(\ref{squeezing}), as
the coupling coefficient between pump and TEM$_{01}$ is $0.64$ times
smaller than the coupling between pump and TEM$_{00}$.

\begin{figure}%
 \includegraphics[width=\columnwidth]{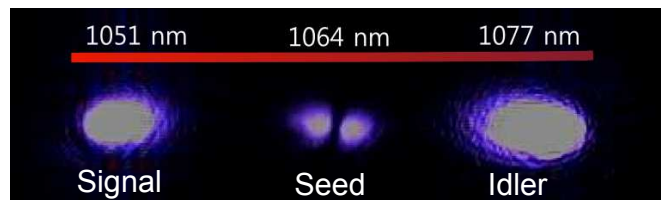}%
 \caption{Wavelength measurement of the signal and idler beams
   emitted in a TEM$_{00}$ mode when the OPO is seeded with a
   TEM$_{10}$ at $1064nm$.}%
 \label{frequencetem00}%
\end{figure}

The lockings of the cavity length and the relative phase being
independent from the OPO, we can investigate the quantum behaviour of
the output modes through the oscillation threshold. In our
case, when locked on the TEM$_{10}$ at $1064nm$, the transverse
mode emitted by the OPO above threshold depends on the mode-matching
of the pump. We chose to have it emit a couple of frequency non-degenerate signal and idler TEM$_{00}$. This is possible
thanks to the very large phase-matching curves of the PPKTP
crystal. Using a diffraction grating we measured the frequency of the
TEM$_{00}$ signal and idler modes and found $\lambda=1051nm$
and $1077nm$ as shown on Fig. \ref{frequencetem00}. This value
agrees with the theoretical prediction taking into account the
dispersion inside the PPKTP crystal \cite{PPKTP}, and the value of the
Gouy Phase of the gaussian modes inside the
cavity.

\begin{figure}%
 \includegraphics[width=\columnwidth]{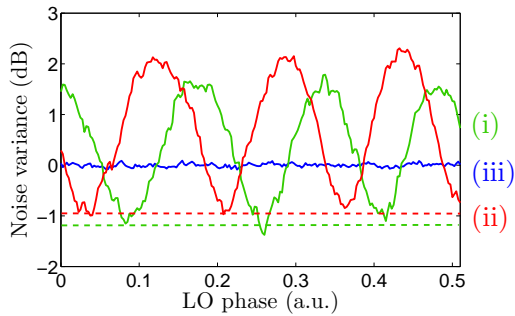}
 \caption{Squeezing measured on the TEM$_{01}$ vacuum mode at the
   oscillation threshold of the OPO. The green curve (i) is just
   below the threshold, whereas the red curve (ii) is just above. We
   observe in both cases $1\pm 0.2dB$ of squeezing. The
   anti-squeezing on the orthogonal quadrature is increased at the
   threshold, from $1.7\pm 0.2dB$ to $2\pm 0.2dB$.}%
 \label{presseuil}%
\end{figure}

\begin{figure}%
 \includegraphics[width=\columnwidth]{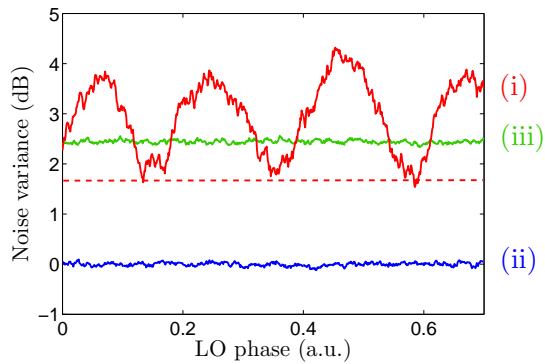}
 \caption{Squeezing measured on the TEM$_{10}$ seed mode above the
   oscillation threshold of the OPO. The red curve (i) represents the
   measured noise, the blue curve (ii) the noise of the local
   oscillator, and the green curve (iii) represents the shot noise
   defined as a corrected value of the shot noise taking into account
   the bright emission. For several pump powers, we observe $1\pm
     0.3dB$ of squeezing and $2\pm 0.3dB$ of anti-squeezing.}%
 \label{dessusseuil}%
\end{figure}

The same homodyne detection as described before is used to investigate
the multimode behavior of the output of the OPO above threshold. The
results follow the predictions of the theoretical part of the present
paper and we observed multimode squeezing with an amount of squeezing
independaent from the pump intensity. The results are first shown on
Fig. \ref{presseuil} when crossing the oscillation threshold, and on
Fig. \ref{dessusseuil} further above when the TEM$_{00}$ emission is
bright, around $3mW$ of power. For different pump powers, we
measured $1dB$ of squeezing and $2dB$ of anti-squeezing on
the two transverse modes, which is the same as below the
threshold. When the TEM$_{00}$ emission is bright, we have to correct
the value of the shot noise measured with the local oscillator alone,
because the homodyne detection photodiodes measures the bright
emission as well as the low-power TEM$_{10}$ and the vacuum
TEM$_{01}$.

We characterized that the state produced by the OPO above the
oscillation threshold is an intrinsic tri-mode state, as defined by
\cite{Treps}. Using the most simple multimode OPO, we generated and
characterized a beam in which the energy is carried by one mode, and
two orthogonal modes carried non-classical features. This
demonstration of the pump clamping inside OPOs sets a new regime
within easy reach of the experimentalists to produce multimode non-classical states
and for which the squeezing is independent from the pump
power. Moreover, the multimode features of the OPO are preserved above
the oscillation threshold. This device is thus a potential stabilized
source for quantum information protocols in the continuous wave
regime. Highly
multimode operation can be potentially performed using the
synchronously pumped OPO described in \cite{Patera1} or the OPO in a
self-imaging cavity described in \cite{Lopez}.

We would like to thank German de Valcarcel and Hans-A. Bachor for
fruitful discussions. We acknowledge the financial
support of the Future and Emerging Technologies (FET) programme within
the Seventh Framework Programme for Research of the European
Commission, under the FET-Open grant agreement HIDEAS, number
FP7-ICT-221906
\\


\begin{thebibliography}{99}
\bibitem{KimbleWu} L.A. Wu, H.J. Kimble, J.L. Hall and H. Wu,
 Phys. Rev. Lett. {\bf 57}, 2520 (1986)
\bibitem{Heidmann} A. Heidmann, R. J. Horowicz, S. Reynaud, E. Giacobino, and C. Fabre, Phys. Rev. Lett. \textbf{59}, 2555 (1987)
\bibitem{Kasai} K. Kasai, Gao Jiangrui, C. Fabre, Europhysics Lett. \textbf{40}, 25 (1997)
\bibitem{Peng} Z. Y. Ou, S. Pereira, H. J. Kimble, K. C. Peng, Phys. Rev. Lett. \textbf{68}, 3663 (1992)
\bibitem{Villar} A. Coelho, F. Barbosa,1 K. Cassemiro, A. Villar, M. Martinelli, P. Nussenzveig, Sciencexpress, 10.1126/science.1178683 (2009)
\bibitem{Schnabel} H. Vahlbruch, M. Mehmet, N. Lastzka, B. Hage, S. Chelkowski,
A. Franzen, S. Gossler, K. Danzmann, R. Schnabel, Phys. Rev. Lett. \textbf{100}, 033602 (2008)
\bibitem{Solvay} C. Fabre, Physics Reports {\bf 219}, 215 (1992)
\bibitem{Gigi} L. Lugiato, P. Galatola, L. Narducci, Optics Commun. {\bf 76}, 276 (1990)
\bibitem{Valcarcel1} I. Pérez-Arjona, E. Roldan and G.J. Valcarcel,
 Euroohys.  Lett. (2006) {\bf 74} 247 (2006)
\bibitem{Navarrete} C. NavarreteBenlloch, E. Roldan, and G.J. de
 Valcarcel, Phys. Rev. Lett. \textbf{100}, 203601 (2008)
\bibitem{Navarrete2}C. Navarrete-Benlloch, G.J. de Valcarcel and
 E. Roldán, Phys Rev A \textbf{79} 043820 (2009)
\bibitem{Patera1} G.J. de Valcarcel, G. Patera, N. Treps and C. Fabre,
 Phys. Rev. A \textbf{74}, 061801(R) (2006)
\bibitem{Patera2} G. Patera, G.J. de Valcarcel, N. Treps and C. Fabre,
 to be published in Eu. Phys. Journal D (2009)
\bibitem{Lopez}L. Lopez, B. Chalopin, A. Rivière de la Souch\`ere,
 C. Fabre, A. Ma\^itre, and N. Treps, Phys. Rev. A \textbf{80},
 043816 (2009)
\bibitem{Lassen2009} M. Lassen, G. Leuchs, and U.L. Andersen,
 Phys. Rev. Lett. \textbf{102} 163602 (2009)
\bibitem{Janousek} J. Janousek, K. Wagner, J.F. Morizur, N. Treps,
 P.K. Lam, C.C. Harb, and H.A. Bachor, Nature Photonics \textbf{3}
 399 (2009)
\bibitem{Lassen2006} M. Lassen, V. Delaubert, C. Harb, P.K. Lam,
 N. Treps, and H.A. Bachor, JEOS RP \textbf{1} 06003 (2006)
\bibitem{PDH} R.W.P. Drever, J.L. Hall, F.V. Kowalski, J. Hough,
 G.M. Ford, A.J. Munley, and H. Ward, Appl. Phys. B \textbf{31} 97
 (1983)
\bibitem{PPKTP} K. Fradkin, A. Arie, A. Skliar, and G. Rosenman,
 Appl. Phys. lett., \textbf{74} 914 (1999)
\bibitem{Treps}N. Treps, V. Delaubert, A. Ma\^itre, J.M. Courty, and
 C. Fabre, Phys. Rev. A \textbf{71} 013820 (2005)
\end{thebibliography}
\end{document}